# Real-time Virtual Intraoperative CT for Image Guided Surgery

Yangming Li[a], Neeraja Konuthula[b], Ian M. Humphreys[b], Kris Moe[b], Blake Hannaford[c], Randall Bly[d]

[a]Rochester Institute of Technology, RoCALab, Rochester, USA, 14623
[b]University of Washington, Department of Otolaryngology–Head and Neck Surgery, Seattle, USA, 98195
[c]University of Washington, BioRobotics Lab, Seattle, USA, 98195
[b]Seattle Children's Hospital, Seattle, USA, 98105

**Abstract.**
**Purpose:** This paper presents a scheme for generating virtual intraoperative CT scans in order to improve surgical completeness in Endoscopic Sinus Surgeries (ESS).
**Approach:** The work presents three methods, the tip motion-based, the tip trajectory-based, and the instrument based, along with non-parametric smoothing and Gaussian Process Regression, for virtual intraoperative CT generation.
**Results:** The proposed methods studied and compared on ESS performed on cadavers. Surgical results show all three methods improve the Dice Similarity Coefficients $> 86\%$, with F-score $> 92\%$ and precision $> 89.91\%$. The tip trajectory-based method was found to have best performance and reached $96.87\%$ precision in surgical completeness evaluation.
**Conclusions:** This work demonstrated that virtual intraoperative CT scans improves the consistency between the actual surgical scene and the reference model, and improves surgical completeness in ESS. Comparing with actual intraoperative CT scans, the proposed scheme has no impact on existing surgical protocols, does not require extra hardware other than the one is already available in most ESS overcome the high costs, the repeated radiation, and the elongated anesthesia caused by actual intraoperative CTs, and is practical in ESS.

*Y. Li, yangming.li@rit.edu, R. Bly, Randall.Bly@seattlechildrens.org

## 1 Introduction

Intraoperative CT scans are the gold standard for confirming surgical modifications in craniofacial surgery. Despite the impressive progress in intraoperative CT technology, this technology exposes patients to repeated radiation and elongated anesthesia, and increases operating room time and space requirements. These limitations jeopardize the applicability and the adoption of intraoperative CTs, making the risk rather unfavorable for cases such as Endoscopic sinus surgery (ESS).[1–4]

Endoscopic sinus surgery (ESS) operates on the paranasal sinuses for treatment of chronic rhinosinusitis (CRS) and resection of neoplasms.[5–7] ESSs is one of the common surgical procedures, and more than 350,000 ESSs are performed annually in the United States alone.[8] ESSs are technically challenging due to the limited field of view of an endoscope, the confined workspace, and close proximity to critical structures such as the optical nerve.[6,9]

Image Guided surgery (IGS) refers to the intraoperative use of preoperative Computed Tomography (CT) images and commercial navigation equipment to locate surgical instruments within the surgical cavity, much like Global Positioning Systems (GPS) navigate with reference to a map (preoperative CT images are the "map" in IGS).[10] With IGS, the intraoperative postion of an instrument in the surgical site can be cross-referenced to the CT images. Through visualizing the instrument position and the preoperative CT, surgeons can navigate inside the surgical cavities. Due to the technical challenges in ESSs, IGS is used routinely and recommended in statements from both the American Academy of Otolaryngology - Head and Neck Surgery (AAO-HNS) and the American Rhinologic Society (ARS).[11,12]

Despite the wide adoption of IGS in ESSs existing IGS systems use a static set of preoperative CT images, which progressively diverge from the actual anatomies as a surgical procedure progresses.[13] The visual explanation of the divergence problem is shown in Fig. 1a. Despite the rapid development of surgical technology, surgeons need to cognitively align the diverged two-dimensional CT images (Fig. 1a) with a real-time two-dimensional endoscopic video (Fig. 1b). This further increases the technical difficulty. ESS suffers from 28% to 47% revision rate, and this is no solution to bridge the gap from both training and clinical perspectives.[14–25]

The conflict between the clinical need for an intraoperative CT update and the intolerance to existing intraoperative CT techniques inspires the research on virtual intraoperative CTs. The majority of existing efforts on addressing the problem lay on reconstructing anatomical surfaces from endoscopic videos and detecting modifications by comparing the reconstructed surface with the preoperative CT. From the methodology perspective, these existing results can be divided into four categories. The first category of methods is known as Structure from Motion (SfM) in the computer vision field.[26] The key technique in SfM is to triangulate correspondence across two or more images to recover the depth information. This technique has a long history in the computer vision field and is broadly studied for reconstructing anatomical surfaces.[27–30] The second category is called Shape-from Shading (SfS), which, as indicated by the name, reconstruct surfaces through explicitly modeling the reflectance of the object surface under known illumination conditions.[31,32] Because endoscopic illumination is coaxial with imaging axis, this technique is utilized in endoscopic surgeries.[33,34] The third category is Visual Simultaneous Localization and Mapping (vSLAM).[35] vSLAM attracts interests in the surgical field, as although this technique is similar to SfM, vSLAM is more precise because it globally tracks landmarks, and thus can optimize trajectories and landmarks.[36] The fourth category is the popular deep learning technique. Many techniques in the deep learning field were introduced and optimized to estimate depth from endoscopic videos.[37,38] Due to the challenges, such as diversity of patient anatomy, dramatic changes in illumination and scene depth, unpredictable adverse imaging conditions due to mucus and bleeding, the combination of deformation and modification, and nearly no texture on some tissues' surfaces, estimation of depth remains a challenging problem.

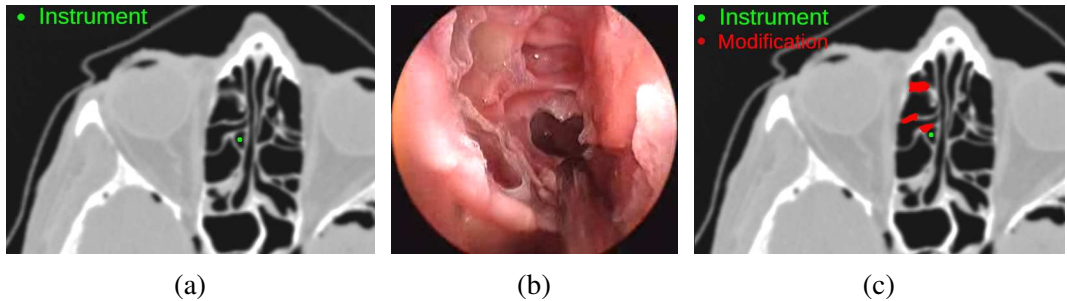

|  (a) | (b) | (c) |

Fig 1: Problem of Existing IGS System. (a) Existing navigation systems depend on preoperative CT, which remains static in the surgery, and causes inconsistency and confusion (an example instrument, indicated by the blue line, overlapped with bony structures). (b) Surgeons can only perceive modifications from endoscopic images and deduce the modification to CT. (c) The proposed research will explicitly detect surgical modification (red area), based on the instrument trajectory and endoscopic video.

This work presents a novel solution to generate virtual intraoperative CT scans for Endoscopic Sinus Surgeries (ESSs) utilizing image navigation data already available in operating rooms. This solution exploits the fact that tissues in ESSs are confined by bony structures and have limited movement and deformation, and uses instrument motion data and preoperative CT scans to generate the virtual CT in real-time. The existing data has limited information, limited positioning reliability, and low sampling rate, and is insufficient for identifying surgical modifications. The proposed solution addresses these limitations with non-parametric filtering and Gaussian Process Regression (GPR). It is passive and has no impact on existing surgical protocols. In summary, this paper has following contributions:

- novel methods are developed to estimate surgical modifications in ESS based on commercial surgical navigation data, which is currently available in ESS;

- virtual intraoperative CT scans are produced based on the estimated imparted surgical tissue change, and the methods neither delay surgeries, nor expose patients to radiation;

- the inconsistency between data from IGS and actual surgery is addressed, which can improve surgical completeness assessment in real time;

- the methods are generally applicable to Endoscopic Sinus and Skull-base Surgeries, and it neither requires extra equipment nor impacts the currently in-use surgical protocols.

The paper is organized as follows: Section 2 studies the characteristics of an example commercial surgical navigation system and introduces three methods to predict surgical change based on instrument motion data. Section 3 verifies the three methods in cadaver experiments, quantitatively assesses the performance with the Dice Similarity Coefficients (DSC), the balanced F-score, the precision, the recall rate, and the Hausdorff Distance, and evaluated the clinical effectiveness of using virtual intraoperative CTs to evaluate surgical completeness, compared to the gold standard of actual intraoperative CTs. Conclusions are drawn in Section 4.

## 2 Method

This paper aims to estimate surgical modifications and update the preoperative CT to reflect the actual surgical scene in a real-time manner, as a virtual intraoperative CT scan. The intuition behind the paper is simple: if a surgical instrument "overlaps" with an anatomical tissue, then the tissue is either removed or deformed. However, commercial surgical navigation systems have a low sampling rate, limited tracking precision, and tracking reliability, and limited information, it is challenging to reliably and precisely estimate surgical tissue removal. We studied modern surgical navigation systems in order to address these problems.

Modern surgical navigation systems for Image Guided Surgeries (IGSs) work as a Global Position System (GPS): satellites are replaced by cameras, infrared sensors, or electromagnetic emitters, and trackers mounted on instruments serve as the "GPS receivers". Through simultaneously tracking a patient and the operating surgical instruments, the geographical information between instruments and the patient is known. This geographical information is projected into the preoperative CT, which is the map in a surgery. Because these systems are designed for positioning surgical instruments in IGSs, they often have a low sampling rate (5∼25 Hz), limited precision (up to 2 millimeters), and only provide location information.[10] More discussion on instrument motion is described in the next subsection.

## 2.1 Data Collection and Statistical Study

The proposed methods aim to provide a ready-to-use technique for ESSs, which only depends on existing FDA-approved and widely used commercial surgical navigation systems in ESS.[11,12] These commercial surgical navigation systems utilize various imaging mechanisms for positioning, but they share the same design goal and have similar characteristics. This work used Medtronic (Minneapolis, MN, USA 55432) StealthStation® S7®, as it is one of the widely adopted commercial surgical navigation systems in USA.[39]

The data collection system was established and the characteristics of motion data were studied. The data collection hardware is the off-the-shelf S7 navigator (Fig. 2). The self-developed software that reads motion data from S7 in real-time is available on a public git repository.[40]

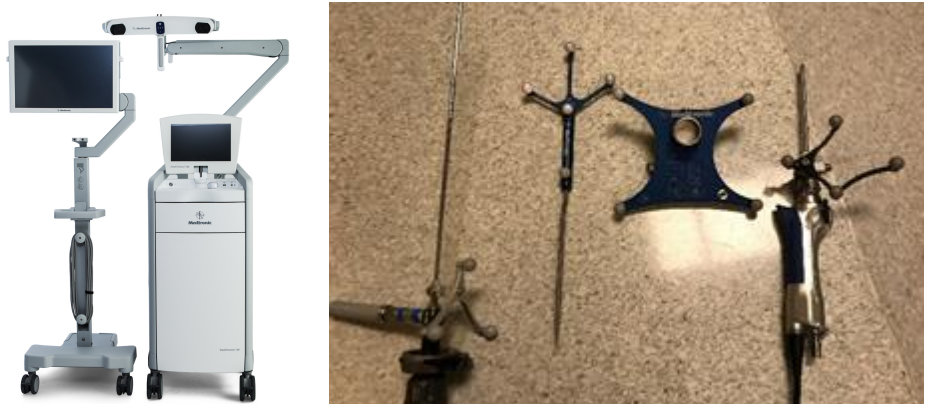

(a) Medtronic S7 Navigation Station.  (b) Surgical motion tracking tools including the patient tracker and instrument trackers. .

Fig 2: System for Data Collection.

The sampling rate and the tracking rate decrease were calculated based on cadaver ESS. To minimize the impact of technical factors on these parameters, the developed software asynchronously communicated with the surgical navigation system. The tracking rate is defined as the ratio between the number of valid samples and the total number of samples. The results from 4 cadavers (about 2 hours long in total) are shown in Table 1. Both the sampling rate and the tracking rate are low in actual ESSs, which indicates data loss and increases the technical challenge on surgical removal estimation.

The localization precision, sampling rate, and tracking rate were analyzed. The localization precision was studied in static experimental setups, in which a high-precision laser scan (error < 0.01 millimeters) of a fixture served as the ground truth (Fig. 3). The fixture was located in different positions inside the field of view of the S7, and the positions of fiducial points on the fixture were measured by the S7. The measured fiducial positions were globally aligned with the ground truth positions to minimize alignment errors.[41] The results showed the standard deviation was 0.37 millimeters.

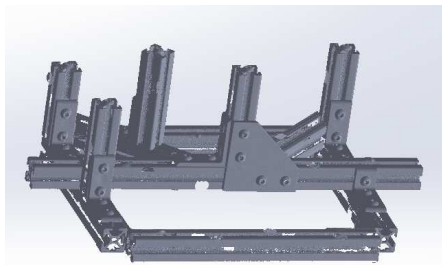

Fig 3: Fixture for Characterizing Tracking Precision.

Table 1: Motion Data Statistics.

| Specimen Number | Sampling Rate (Unit: Hz) | | Tracking Rate (Unit: %) |
|---|---|---|---|
| | High | Low | |
| C1 | 14.11 | 5.68 | 76.72 |
| C2 | 15.02 | 5.42 | 72.78 |
| C3 | 15.47 | 5.53 | 70.15 |
| C4 | 15.58 | 5.53 | 71.61 |

Note: The sampling rate and the tracking rate depend on experimental setup. The numbers above only indicates the sampling rate and the tracking rate in 4 cadaver ESSs.

Based on the characteristics of the motion data, the following three methods were proposed to address the problems of low sampling rate and low tracking rate.

## 2.2 Method 1: Removal From Instrument Tip Motion

For most of the surgical instruments, tissue removal occurs at the tip of instruments. Despite the simplicity of the concept, it is challenging to predict removal due to the localization errors and the insufficient information.

It is challenging to address the insufficient information problem because contact between an instrument tip and an anatomical structure does not necessarily cause modifications.[42] It is difficult to differentiate the motions that cause surgical removal from the motions that only lead to deformation. This paper proposes using "tip density" to improve surgical removal estimation reliability. The tip density is defined as the duration an instrument tip resides within a set of voxels representing an anatomical structure. The assumption behind this is that surgeons are cautious when removing anatomical tissue thus the duration of modification correlate higher than the duration of deformation in ESSs, and the "tip density" is an approximation of the duration. To implement this, the 3-dimensional space was raterized to align with a CT voxels and normalized the duration according to the length of the surgical procedure. The threshold of the duration is determined by the equation below(Eqn. 1a). Any voxel that has a density lower than the cutoff threshold is considered unchanged. In Eqn. 1a, $d$ is the tip density, $v(d)$ is the total numbers of voxels with tip density $d$, $\ddot{v}(d)$ and $\dddot{v}(d)$ are the second and the third-order gradients of $v(d)$, and $th$ is the mean value of $v(d)$.

To address the localization error problem, the localization error was modeled as a Gaussian distributed zero-mean random variable (measured in the previous subsection) with a non-parametric smoothing algorithm to improve localization precision.[43]

$$\arg\min_{d} v(d) = \{d | \dddot{v}(d) > th, \ddot{v}(d) \leq th, d \in \mathbb{N}\} \tag{1a}$$

$$th = \text{mean}(v(d)) \tag{1b}$$

Fig. 4a shows the tip densities as a heat map. A visual representation of the voxels to be deleted from the tip density algorithm is displayed in Fig. 4b.

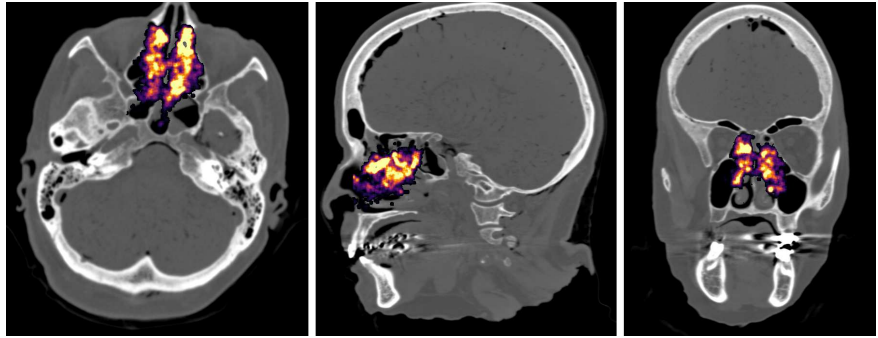

(a) Tip Density Heat Map.

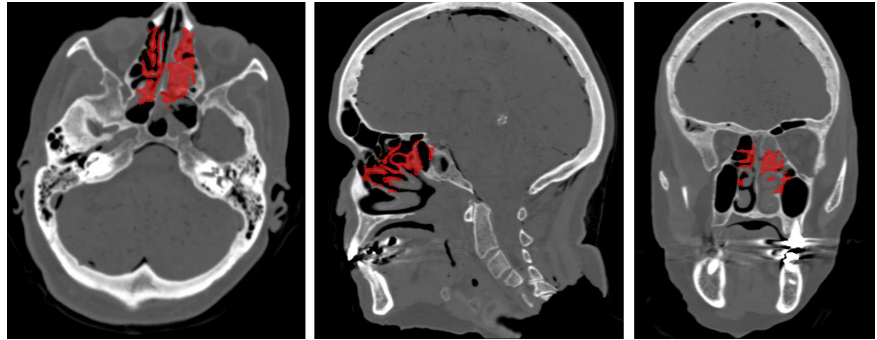

(b) Tip Based Removal Estimation.

Fig 4: Instrument Tip Based Surgical Removal Estimation. The left figure (a) shows the tip density values heat map, where the higher redness, the higher density. The right figure (b) shows the estimated surgical removal based on tip density. Tip density roughly reflects the duration an instrument tip stays in a CT voxel and is used to differentiate actual surgical removal from tissue deformation.

## 2.3 Method 2: Removal From Instrument Tip Trajectory

Based on the continuity of human motions, this method addresses the low sampling rate and the low tracking rate problems by estimating the full instrument tip trajectories. Although there are a number of studies on the human hand and finger motion patterns, this method adopts Gaussian Process Regression (GPR) to learn the motion model from the available motion data. In the estimation of the trajectory, the instrument motion was first projected into the endoscope coordinate to minimize the disturbance from unnecessary motion data. The intuition behind this hypothesis is in Image Guided Surgeries, "necessary motion" only occurs while the instrument is visible from the endoscope. The endoscopic coordinate is defined as *the origin is the center of the endoscope tip, the X-positive is the vector starting from the origin and pointing to the hand-held end of the endoscope, and the X-Y plane is formed by the X-axis and the tracker plane norm.*

GPR is used to learn the motion model from the projected data. GPR is a form of kernel based data-driven learning technique.[44] GPR is similar to Bayesian Linear Regression (BLR) in the sense that they estimate the expected output $Y_*$ given input $X_*$ with respect to the training set $(X, Y)$ under the Bayesian framework as: $p(Y_*|X_*, X, Y)$, where we denote the relationship between the input $X$ and output $Y$ as: $Y = f(X) + \omega, \omega \sim \mathcal{N}(0, \sigma_n^2)$. GPR is superior to Bayesian Learning Regression in this application because 1) it is a nonparametric learning technique so it does not explicitly assume the surgical motion model, and 2) it is efficient in data-dense applications.[45] In this work, the observation is defined as: $X = \{x_t, y_t, z_t, \Delta t\}$ and the estimation is defined as:

$Y = \{x_{t+1}, y_{t+1}, z_{t+1}\}$. All motion data are used to train the GPR, which leads to covariance:

$$K = \begin{pmatrix} k(X_1, X_1) & \cdots & k(X_1, X_n) \\ \vdots & \ddots & \vdots \\ k(X_n, X_1) & \cdots & k(X_n, X_n) \end{pmatrix}$$

, and the estimated trajectory $Y_*$ is:

$$Y_*|Y \sim \mathcal{N}(K_* K^{-1} Y, K_{**} - K_* K^{-1} K_*^T) \tag{2a}$$

$$K_* = (k(X_*, X_1) \cdots k(X_*, X_n))$$

$$K_{**} = k(X_*, X_*)$$

Notice that because GPR is a smoothing filter, the instrument raw locations are used in model learning. The estimated trajectory is rasterized based on the corresponding preoperative CT scan's resolution to generate tip densities (Fig. 5).

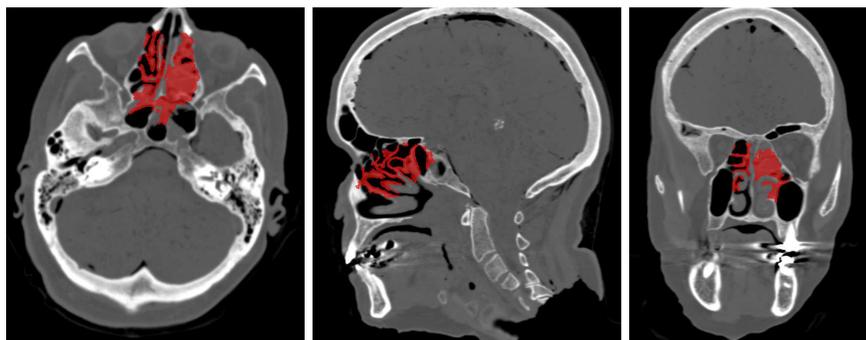

Fig 5: Instrument Tip Trajectory Based Surgical Removal Estimation. Tip trajectory was estimated based on tip locations and GPR, which address the localization error, the low sampling rate, and the low tracking rate problem.

## 2.4 Method 3: Removal From Instrument Motion

Through calibrating the tracking system, the full body of surgical instruments and endoscopes can be localized. Then the body is rasterized based on the corresponding preoperative CT scan resolution to generate tip densities (Fig. 6).

While the previous two methods hypothesize that surgical tissue removal occurs at instrument tips, the instrument and the endoscope body can also be used to detect modifications, because the existence of a rigid body indicate anatomical tissues are either moved or removed.

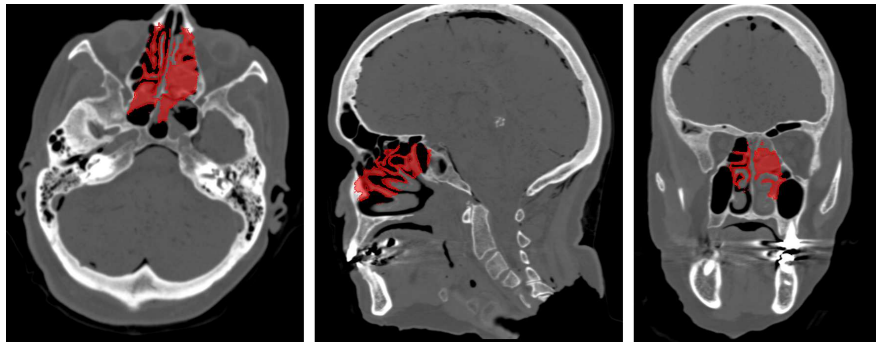

Fig 6: Instrument Body Based Surgical Removal Estimation.

## 2.5 Virtual Intraoperative CT

Intuitively, the virtual intraoperative CT is to subtract the estimated removals from the preoperative CT. For implementation, the estimated removals are used as masks, and all preoperative CT voxels inside the masks are replaced by the minimum intensity voxel inside the mask, to generate the virtual intraoperative CT (Fig. 11).

## 3 Experiment

The generated virtual intraoperative CTs are tested for accuracy on the effectiveness of confirming surgical tissue removal as compared to ground-truth intraoperative CT scan and the completion of ESS. Two different types of evaluation are used in this work. The first one is to quantitatively measure the precision of the estimated removal comparing the virtual CT to the actual CT, based on the following metrics, the Dice Similarity Coefficient, Precision, Recall Rate, and Hausdorff Distance.

Both types of evaluations are conducted on cadaver experiments. Although the proposed methods have no impact on existing surgical protocols in ESSs and require no extra equipment, we only conducted experiments on cadavers so that the surgical procedure can have a range of completeness. The adopted surgical navigation system is shown in Fig. 2, in which Fig. 2a shows Medtronic S7® and 2b shows the tracked endoscope and the microdebrider.

### 3.1 Ground-truth

The ground-truth for the quantitative removal estimation precision is true surgical removal as determined on intraoperative CT scan. The paranasal sinuses, middle turbinates, and septums were manually labeled as the regions of interest. The metrics for these regions of interest shows the performance of the proposed methods as these regions define the boundary of ESS.

Three attending surgeons were asked to evaluate the completeness of the experimental operations. In ESSs, this process is objective, because surgical completeness is evaluated by the openings of cells. The evaluation questionnaire adopted in this work is shown in Table 3.

### 3.2 Virtual Intraoperative CT Precision

As explained in Subsection 3.1, the ground-truth removal is determined by comparing an intraoperative CT with the corresponding preoperative CT. The two CT scans are aligned (Fig. 7), then the threshold is set to -800 to exclude fluid and air from the comparison. An example -800 mask is shown in Fig. 8.

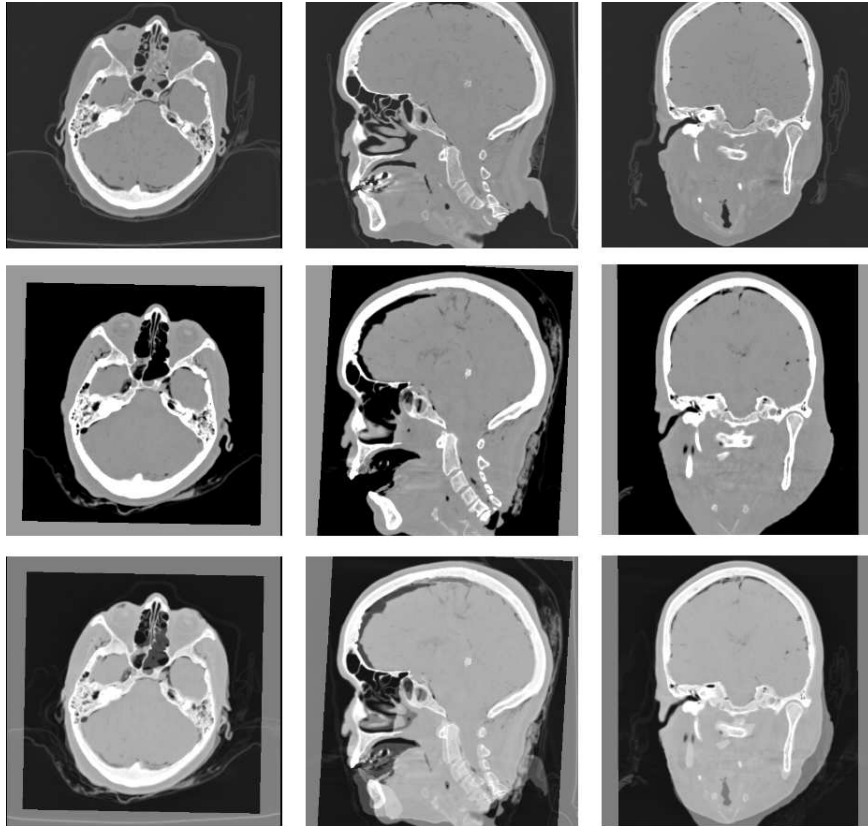

Fig 7: CT Alignment. Preoperative CTs and intraoperative CTs are manually aligned to generate the ground truth for quantitatively measuring the surgical removal estimation precision.

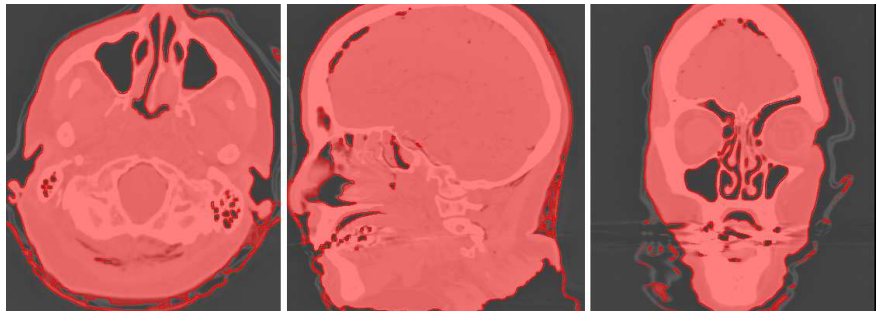

Fig 8: Using threshold to exclude fluid and air from the comparison. -800 was used to the threshold given the CT scanner used in this work.

An example of the segmented regions of interest is shown in Fig.9.

The Dice Similarity Coefficients (DSC), the balanced F-score, the precision, the recall rate, and the Hausdorff Distance are often adopted for demonstrating the similarity of two regions.[46] DSC is defined as $DSC = \dfrac{2|X \cap Y|}{|X|+|Y|}$, where $X$ and $Y$ are the matches between the estimated results and the ground-truth results. The precision, $p$, is defined as: $p = p_t/(p_t + p_f)$, where $p_t$ is the true positives, and $p_f$ is the false positives. The recall rate, $r$, is defined as: $r = p_t/(p_t + n_f)$, where $n_f$ is the false negatives. The balanced F-score, $F$, is defined as: $F = 2 \cdot \dfrac{p \cdot r}{p + r}$. The metrics for the generated virtual intraoperative CTs with respect to the true intraoperative CT are shown in Table 4. Due to the deformation of tissues and other errors, the scores are low if the entire CT volume is compared.

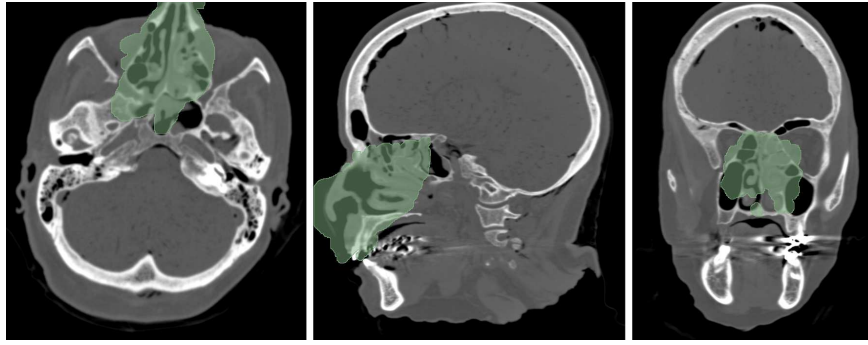

Fig 9: Sinus Segmentation. Sinus Region is manually segmented, in order to eliminate the regions irrelevant to ESS.

we compare the entire head. The regions of interest (ROI) have higher scores than the head mainly because the ROIs are confined by bony structures and have less deformation. The virtual intraoperative CTs have higher scores as they much better reflect the actual surgical scene (Table 4, Row 5∼7 v.s. Row 3 ), comparing with the diverged preoperative CTs.

*3.3 Clinical Completeness Evaluation*

The gold standard for the performance of intraoperative CT scans is whether or not they are sufficient for confirming surgical completeness in ESS. This work assessed the clinical utility of the generated virtual intraoperative CT scans on cadaver experiments. Four cadavers were divided into (by left and right) eight groups of experiments. Surgeons performed classical ESS operations to varying degrees of completeness, including maxillary antrostomy, ethmoidectomy, frontal sinusotomy and sphenoidotomy. Each operation is randomly assigned to be completed, partial, or unoperated, to eliminate the bias of results. The generated virtual intraoperative CT scans and the actual intraoperative CTs are shuffled and anonymized, and three attending surgeons graded the surgeries based on the questionnaire designed for ESSs (Table. 3).

Table 2: Removal Estimation Precision Comparison. The removal estimations from the tip based estimation (Tip), the tip trajectory based estimation (Tip Trajectory), and the instrument body (Body) were quantitatively evaluated with DSC, F-score, Precision, Recall rate, and Hausdorff distance. Metrics are also calculated for the overall CT alignment (Head), Region of Interests (ROI), and the Regions out of Interest (None ROI), as a reference. The results show the mean value $\pm$ the standard deviation.

|  | DSC (Unit: %) | F-score (Unit: %) | Precision (Unit: %) | Recall (Unit: %) | Hausdorff Distance (Unit: mm) |
|---|---|---|---|---|---|
| Head | 79.75±1.17 | 67.59±0.36 | 75.22±0.66 | 61.82±0.63 | 9.77±2.97 |
| ROI | 83.65±0.56 | 81.49±1.78 | 89.28±0.94 | 75.22±2.47 | 1.14±0.98 |
| None ROI | 79.43±1.33 | 66.36±0.52 | 74.00±0.77 | 60.69±0.91 | 9.91±2.71 |
| Tip | 88.50±0.11 | 93.65±0.10 | 89.91±0.08 | 77.72±0.12 | 2.44±1.67 |
| Tip Trajectory | 87.99±0.10 | 93.25±0.05 | 90.80±0.02 | 85.85±0.08 | 1.95±0.83 |
| Body | 86.09±0.09 | 92.05±0.07 | 91.32±0.03 | 82.80±0.08 | 2.93±1.67 |

Table 3: Questionnaire for Surgical Completeness Evaluation in Endoscopic Sinus Surgeries.

|  |  | Fully Opened | Partially Opened | Not Opened |
|---|---|---|---|---|
| Maxillary Sinus | Left | uncinate completely removed | partial uncinate remains | not opened |
|  | Right |  |  |  |
| Anterior Ethmoid | Left | all septations removed | septations remain | multiple unopened cells remain |
|  | Right |  |  |  |
| Posterior Ethmoid | Left |  |  |  |
|  | Right |  |  |  |
| Sphenoid | Left | opened widely for the indication of chronic sinusitis | partially opened | not opened |
|  | Right |  |  |  |

The evaluation results are shown in Fig.10. In the figure, the red stars indicate the completeness evaluation based on the actual intraoperative CT scans, and the green circles, the blue squares and the black diamonds denote the completeness evaluated by the virtual intraoperative CT scans generated by the three proposed methods. The precision is evaluated in Table 4. In the table, the precision is defined by the distance between the two evaluation over the largest distance. The distance is numerically defined as: 0.5 for classifying unopened as partially opened, or partially opened as unopened or fully opened, or fully opened as partially opened; 1 for classifying unopened as fully opened, or fully opened as unopened. Based on the definition, the largest distance is 1. From Fig. 10 and Table 4 it is clear that although all three methods have good performance, the tip trajectory based virtual intraoperative CTs has the best performance and reaches $96.87\%$ precision.

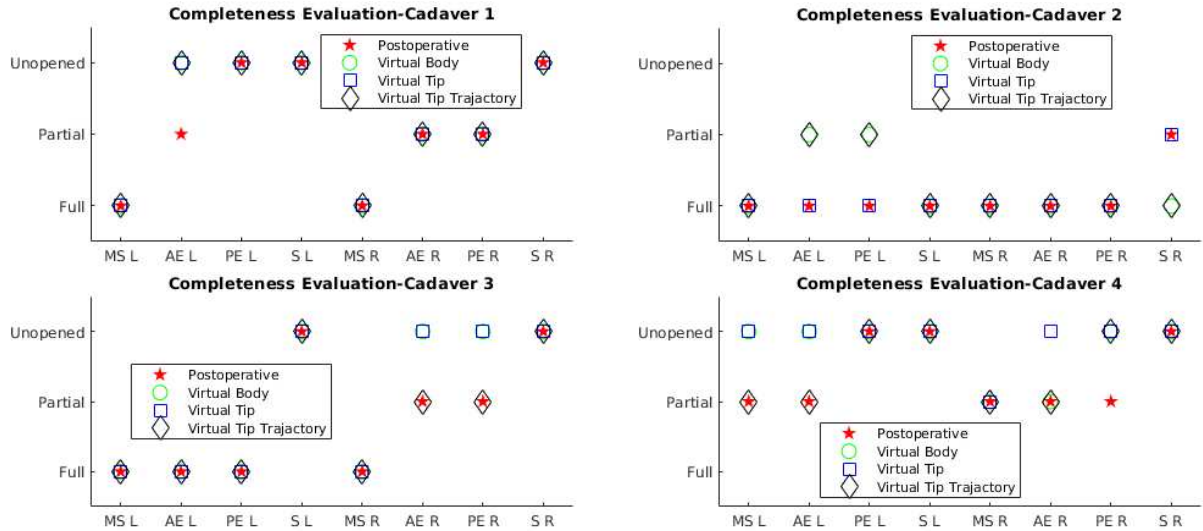

Fig 10: Surgical Completeness Evaluation. Surgical completeness was blindly evaluated based on the virtual CTs and the intraoperative CTs. "MS L" denotes Maxillary Sinus Left, "MS R" denotes Maxillary Sinus Right, "AE L" denotes Anterior Ethmoid Left, "AE R" denotes Anterior Ethmoid Right, "PE L" denotes Posterior Ethmoid Left, "PE R" denotes Posterior Ethmoid Right, "S L" denotes Sphenoid Left, and "S R" denotes Sphenoid Right.

Table 4: Completeness Evaluation Precision Comparison.

|  | MS L | AE L | PE L | S L | MS R | AE R | PE R | S R | **Overall** |
|---|---|---|---|---|---|---|---|---|---|
| Virtual Body | 87.5 | 62.5 | 87.5 | 100 | 100 | 87.5 | 75 | 87.5 | **85.94** |
| Virtual Tip | 87.5 | 62.5 | 87.5 | 100 | 100 | 75 | 75 | 87.5 | **84.38** |
| Virtual Tip Trajectory | 100 | 87.5 | 100 | 100 | 100 | 100 | 87.5 | 100 | **96.87** |

An example of the virtual CT and actual CT comparison is shown in Fig. 11 for visual qualitative comparison.

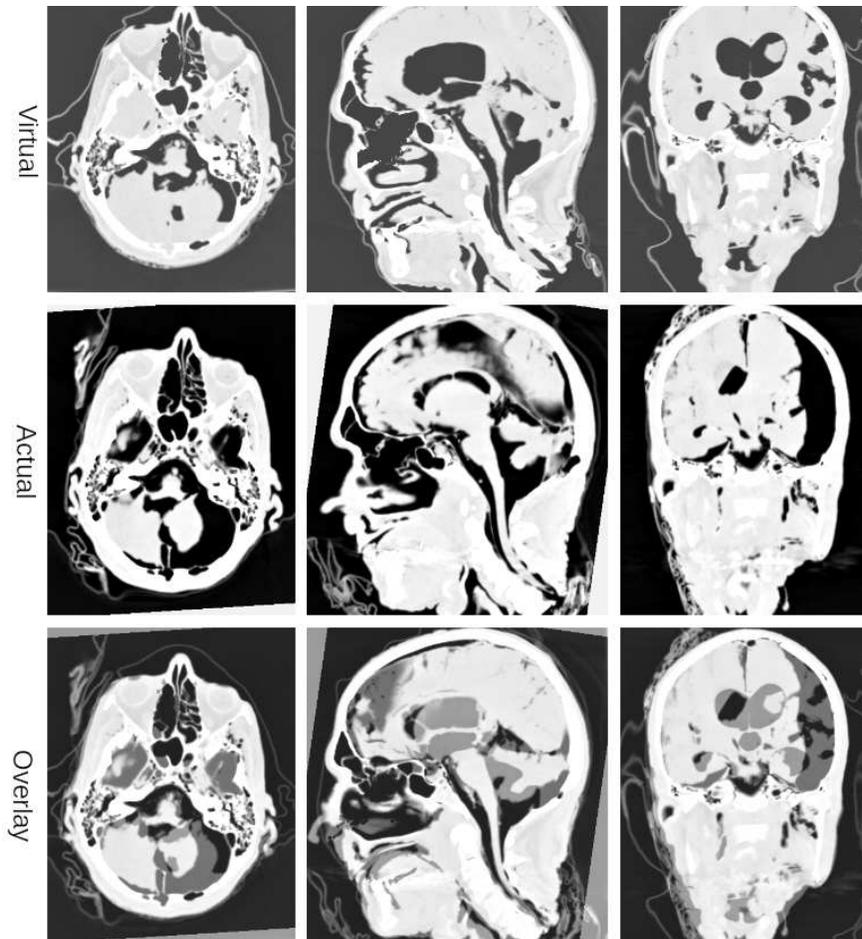

Fig 11: Virtual Intraoperative CT. The bottom row shows the comparison of virtual intraoperative CT (top row) overlaid on the intraoperative CT (middle row).

## 4  Conclusion

Intraoperative CTs are the gold standard for confirming surgical completeness in operating rooms. However, this technology is not practical in many common surgeries, such as Endoscopic Sinus Surgeries (ESSs), due to the disadvantages of increased time, cost and radiation exposure. This work presents a virtual intraoperative CT generation scheme to introduce intraoperative CT into ESSs without the disadvantages. The proposed scheme works with existing commercial surgical navigation systems in use clinically, which is standard of care in most ESSs. The proposed methods were applied to cadaver ESSs and the experimental results demonstrated that although the proposed methods are simple and reliable, they reach $96.87\%$ on surgical completeness prediction precision.